\documentclass[lettersize,journal]{IEEEtran}
\usepackage[T1]{fontenc}
\usepackage{amsmath,amsfonts}
\usepackage{algorithmic}
\usepackage{algorithm}
\usepackage{array}
\usepackage{textcomp}
\usepackage{stfloats}
\usepackage{url}
\usepackage{verbatim}
\usepackage{graphicx}
\usepackage{svg}
\usepackage{xcolor}
\usepackage{cite}
\usepackage{subcaption}
\begin{document}
\title{Modeling Plant Action Potentials under Photoperiod Stress via Hodgkin-Huxley Dynamics}

\author{
    \author{Imen~Bekkari,~Maurizio~Magarini,~\IEEEmembership{Member,~IEEE,}~and~Hamdan~Awan,~\IEEEmembership{Member,~IEEE}%
    }
}

\maketitle

\begin{abstract}

Plants exhibit dynamic bioelectric properties that facilitate information transfer across tissues. This study investigates action potentials (APs) in \textit{Nicotiana tabacum} (tobacco) recorded within a custom-designed growth chamber using a biosignal amplifier and environmental sensors. Consistent light- and dark-induced APs were observed during photoperiod transitions under controlled 12-hour artificial illumination cycles. To understand these bioelectric responses, a mathematical model based on the Hodgkin-Huxley (HH) one is used.  
Electrophysiological measurements from \textit{Solanum lycopersicum} (tomato) revealed that under natural light conditions, only light-induced APs are observed, while light- and dark-induced APs coupled dynamics is exclusively elicited during rapid transitions in artificial photoperiods.
These distinct phenomena are characterized as prolonged oscillatory climatic engagement (POCE) and 
nimble environmental transition oscillation (NETO), respectively. The model successfully reproduces the key features in both frameworks while maintaining computational efficiency through voltage-independent gating kinetics.
\end{abstract}


\section{Introduction}
\label{sec:introduction}
\IEEEPARstart{P}{lants} continuously sense and respond to environmental stimuli through sophisticated signaling networks that integrate electrical, chemical, and hydraulic pathways~\cite{sukhov16, akyildiz08}. 
Electrical signals provide rapid, long-distance communication that operates in concert with slower chemical transport to coordinate defense responses, regulate photosynthesis, and modulate growth~\cite{sukhov16, mudrilov21}. Action potentials (APs), variation potentials (VPs), and system potentials (SPs) represent distinct classes of electrical events characterized by different propagation velocities, amplitudes, and physiological triggers~\cite{awan19,sukhov11}. These electrical waves are intrinsically coupled to transient elevations in cytosolic calcium ($\mathrm{Ca}^{2+}$) and reactive oxygen species (ROS), forming mutually reinforcing signaling modules that propagate through vascular tissues at velocities reaching millimeters per second~\cite{edel17}. The ionic basis of plant APs involves coordinated fluxes of potassium ($\mathrm{K}^+$), chloride ($\mathrm{Cl}^-$), $\mathrm{Ca}^{2+}$ and protons ($\mathrm{H}^+$) across the plasma membrane, mediated by voltage-gated channels and electrogenic pumps~\cite{sukhov09}. 
Recent advances in molecular communication (MC) theory provide a rigorous framework for analyzing plant signaling from an information-theoretic perspective~\cite{kuran21, magarini20}. By abstracting biological processes into encoder-channel-decoder architectures, MC enables quantitative evaluation of information transfer capacity, channel impulse responses, and optimal detection strategies. While this approach has proven valuable for understanding intercellular signaling in synthetic biology and engineered cellular networks~\cite{akyildiz08}, its application to native plant communication systems remains largely unexplored. Integrating MC theory with plant electrophysiology enables quantification of how electrical excitation drives molecular emission, how reaction-diffusion dynamics shape information propagation through vascular tissues, and how downstream cells decode multiplexed signals into coordinated physiological actions.

This paper presents an integrated computational and experimental framework to characterize plant electrophysiological responses under photoperiod stress. The framework extends the established Hodgkin-Huxley (HH) formalism to capture plant APs kinetics, with parameters derived from the literature~\cite{sukhov09,yao23}.
Although light-induced APs are known to occur during changes in illumination~\cite{awan21}, the underlying mechanisms remain incompletely characterized. This knowledge gap  restricts the potential use of bioelectric signals as non-invasive indicators of plant physiological status. Understanding these electrophysiological dynamics is relevant to crop resilience because APs trigger downstream physiological responses, including modulation of photosynthesis, stomatal regulation, and systemic defense gene expression~\cite{sukhov16,mudrilov21}. 
\IEEEpubidadjcol
Accurate characterization and modeling of light-induced APs could therefore enable real-time monitoring of plant stress states and inform adaptive management strategies.

In this work, two distinct electrophysiological phenomena emerge under different illumination regimes. When subjected to abrupt light-dark transitions, tobacco plants generate both light- and dark-induced APs, constituting a bidirectional response pattern herein termed nimble environmental transition oscillation (NETO). In contrast, tomato plants cultivated under natural photoperiodic conditions, where irradiance increases progressively, exhibit APs in response to light onset, with no corresponding electrophysiological activity at dusk. This unidirectional pattern is termed prolonged oscillatory climatic engagement (POCE). Moreover, the light-induced APs differ markedly between these two regimes.
We propose that these differences result from variations in the illumination setup rather than from species-specific factors alone. An HH-based model is used to test this hypothesis by reproducing the distinct waveform characteristics observed under both NETO and POCE conditions.
An additional methodological consideration concerns the use of constant gating rate parameters. Because the recordings span multiple days and the primary aim is model-based characterization of inter-condition variability, this approach reduces parameter dimensionality when fitting nonlinear rate functions to slow AP waveforms.

\subsection{Paper's Organization}
The remainder of this paper is organized as follows. Section~\ref{sec:stateoftheart} reviews the state of the art in plant electrical signaling, MC frameworks, and HH-based modeling for plant APs. The system model is described in Sec.~\ref{sec:system_model}, including the growbox design and electrophysiological data acquisition. Section~\ref{sec:data_analysis} presents the data analysis procedures and feature extraction methods applied to both experimental datasets. Experimental observations and model-fitting results for NETO and POCE responses are reported in Sec.~\ref{sec:results}. Section~\ref{sec:discussion} discusses the physiological implications of the observed waveform differences and the model limitations. Finally, Sec.~\ref{sec:conclusion} concludes the paper and outlines directions for future work.
\begin{figure}[!t]
    \centering
    \includegraphics[width=0.9\linewidth]{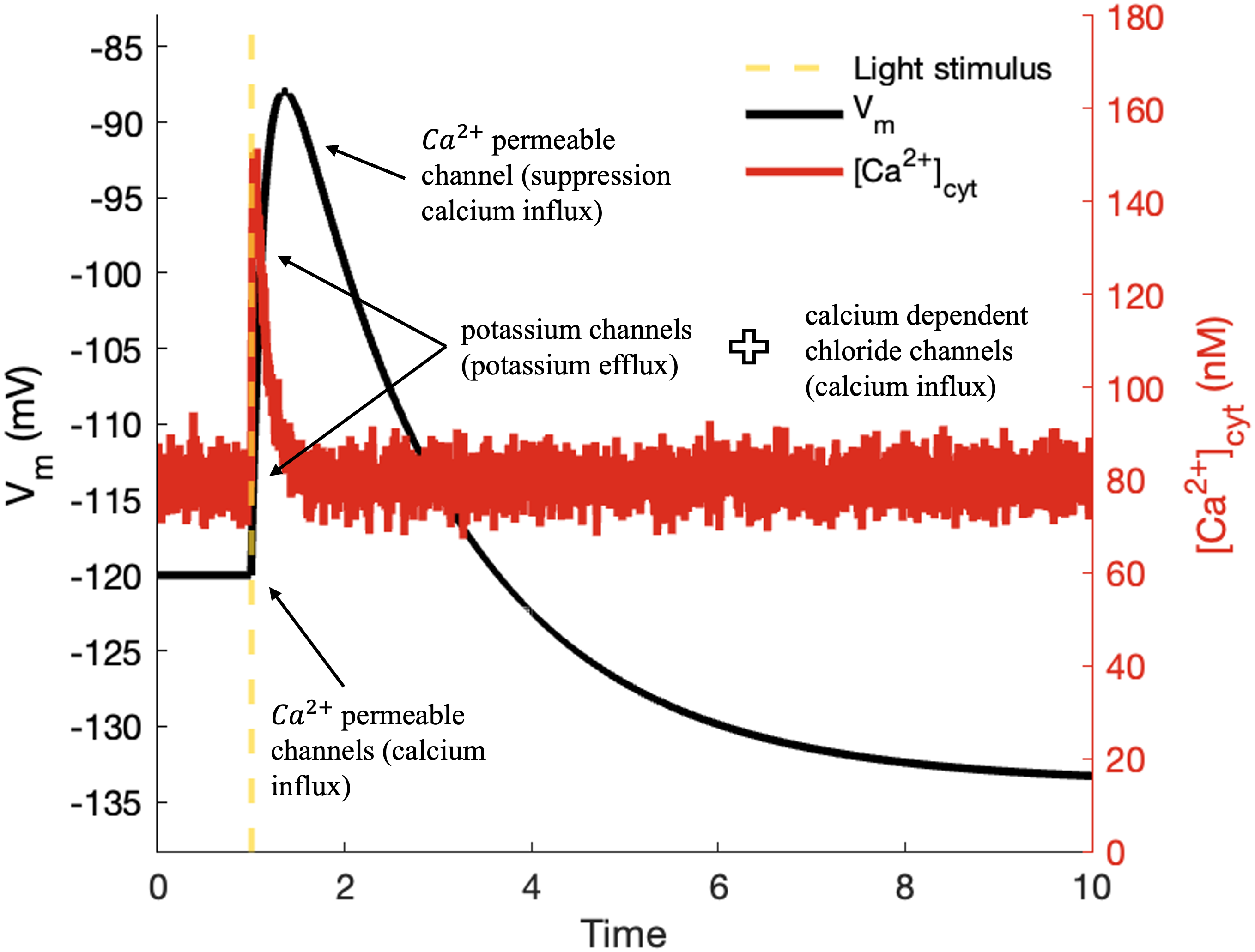}
     \caption{Light-induced AP and $\mathrm{Ca}^{2+}$ response. Black lines indicate the AP, and
    red lines indicate the $\mathrm{Ca}^{2+}$ dynamics. $V_{\mathrm{m}}$ denotes the membrane
    potential. $\mathrm{Ca}^{2+}_{\mathrm{cyt}}$ denotes the cytosolic calcium concentration.
    The yellow dashed line indicates the light onset~\cite{mudrilov21}.}
    \label{fig:light-induced ER}
\end{figure}

\section{State of the Art}
\label{sec:stateoftheart}

This section reviews relevant background.
Section~\ref{sec:electrical_signaling} covers plant electrical signaling and its ionic basis. Section~\ref{sec:MolecularCommunication} introduces the MC framework and its application to plant communication. Section~\ref{sec:HH} describes HH-based models for plant AP generation.

\subsection{Electrical Signaling}
\label{sec:electrical_signaling}
Electrical signaling provides plants with a rapid, system-level coordination layer that operates
alongside chemical and hydraulic pathways to couple local perception of stimuli with distal
physiological responses. Plant cells maintain a large, negative plasma-membrane potential established by the $\mathrm{H}^+$-ATPase and selective ion fluxes
($\mathrm{K}^+$, $\mathrm{Ca}^{2+}$, $\mathrm{Cl}^-$, $\mathrm{H}^+$)~\cite{edel17}.
Perturbations to ion transport by environmental or biotic stimuli evoke transient deviations
(depolarizations or hyperpolarizations) that scale from local microdomains to organ- and
whole-plant signals~\cite{choi16}. At the tissue scale, long-distance signaling integrates three tightly
coupled modules: electrical potentials, cytosolic $\mathrm{Ca}^{2+}$ waves, and apoplastic ROS
waves~\cite{gilroy14}.

\begin{figure}[!t]
    \centering
    \includegraphics[width=0.9\linewidth]{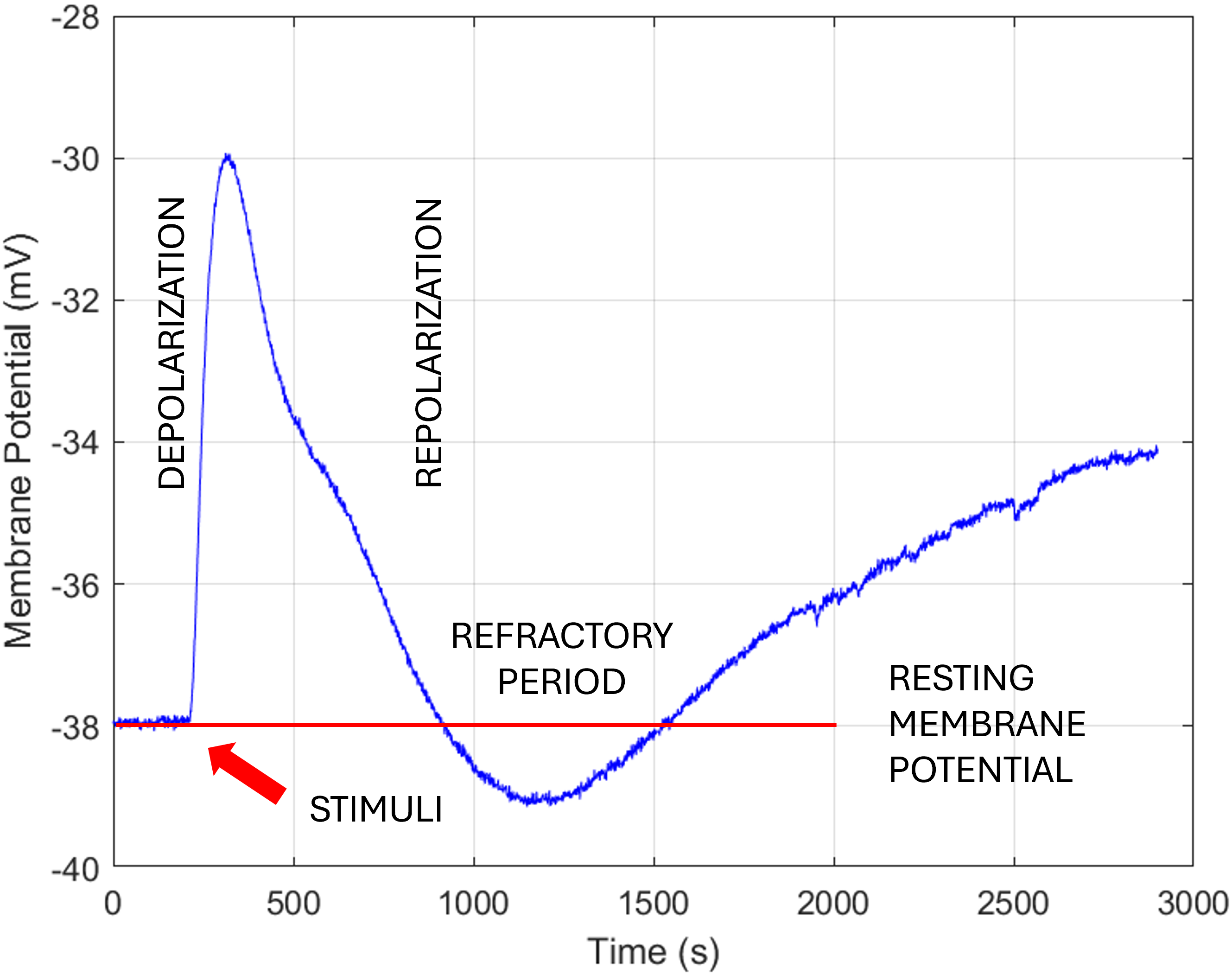}
    \caption{Light-induced AP within a NETO response, recorded from tobacco at the onset of
    artificial light in the Agrowbox dataset.}
    \label{fig:NETO_ER_light}
\end{figure}

APs are all-or-nothing depolarizations evoked by non-damaging stimuli such as touch, light
steps, or cold shock, exhibiting a refractory period and propagating with relatively constant
amplitude and velocity~\cite{choi16, gilroy14}. VPs are graded, longer-lasting depolarizations
elicited by damaging stimuli (e.g., wounding, burning, abrupt thermal shocks), with amplitude
and speed that decay with distance~\cite{li21}. SPs are long-distance, apoplast-associated hyperpolarizations/depolarizations that often accompany AP/VP trains during systemic responses~\cite{garcia21}.

Rapid long-distance signaling arises from tight coupling between $\mathrm{Ca}^{2+}$ dynamics,
ROS, and membrane voltage (Fig.~\ref{fig:light-induced ER}). Local stimuli can launch self-propagating ROS and
$\mathrm{Ca}^{2+}$ waves that prime distal organs for acclimatization~\cite{choi16}. These waves are mutually
reinforcing since $\mathrm{Ca}^{2+}$ elevations activate ROS production, and ROS promote further
$\mathrm{Ca}^{2+}$ entry, creating positive feedback. In the vasculature, fast electrical
changes travel over long distances and are reduced when glutamate receptor-like channels are
disrupted, underscoring the central role of $\mathrm{Ca}^{2+}$ influx in generating plant
electrical signals~\cite{gilroy14}.

\begin{figure}[!t]
    \centering
    \includegraphics[width=0.9\linewidth]{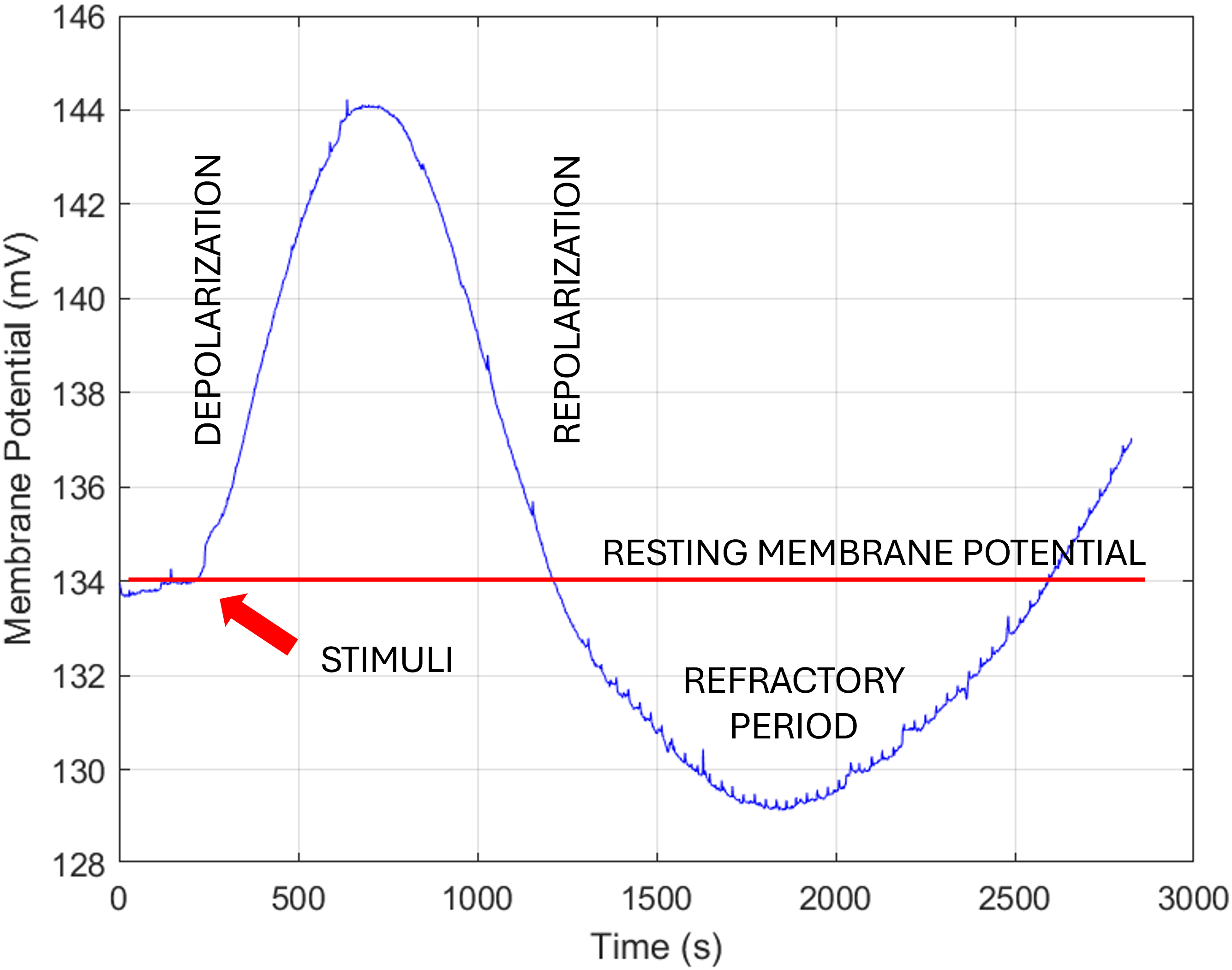}
    \caption{Light-induced AP within a POCE response, recorded from tomatoes at natural sunrise
    (Vivent SA~\cite{vivent}).}
    \label{fig:POCE_ER_light}
\end{figure}

In the canonical AP sequence, voltage- or stimulus-gated $\mathrm{Ca}^{2+}$ entry initiates
depolarization (Figs.~\ref{fig:NETO_ER_light} and~\ref{fig:POCE_ER_light}), activating
$\mathrm{Ca}^{2+}$-dependent anion ($\mathrm{Cl}^-$) efflux. Subsequent opening of
outward-rectifying $\mathrm{K}^+$ channels repolarizes the membrane, with the
$\mathrm{H}^+$-ATPase restoring ionic homeostasis~\cite{sukhov09}.
The physiological impact is broad as ROS/
$\mathrm{Ca}^{2+}$ waves regulate photosynthesis, stomatal conductance, and systemic defense
gene expression (e.g., PIN2, jasmonic acid (JA)/abscisic acid (ABA) components)~\cite{lee23}.
From a sensing perspective, plant electrical
activity can be measured using extracellular electrodes, optical reporters, or ion-selective microelectrodes~\cite{li21, choi16, vivent}. Emerging machine-learning (ML) pipelines classify
stress modalities from field-acquired electrical time series, enabling real-time physiological state estimation~\cite{garcia21, li21}. 

\subsubsection{Electrical Reaction to Light}
Light steps evoke rapid electrical responses governing phototropism and photosynthetic
acclimation. A common waveform under step illumination is a brief depolarization followed by a
transition to a more negative steady $V_{\mathrm{m}}$, reflecting $\mathrm{Ca}^{2+}$-dependent
anion depolarization and subsequent $\mathrm{H}^+$-ATPase-dominated recovery~\cite{koselski08}
(Fig.~\ref{fig:light-induced ER}). In \textit{Arabidopsis} and other angiosperms, APs propagate via the phloem at mm\,s$^{-1}$, while excess light launches slower VP/SP-like waves that co-propagate with $\mathrm{Ca}^{2+}$/ROS signals and transiently reprogram photosynthesis and gas exchange~\cite{favre07, szechynska17}. At canopy and
whole-plant scales, diel light cycles imprint robust day-night patterns in trunk electrical
potentials, highlighting opportunities for continuous electrical sensing in the
field~\cite{oyarce10}.

\begin{figure}[!t]
    \centering
    \includegraphics[width=0.9\linewidth]{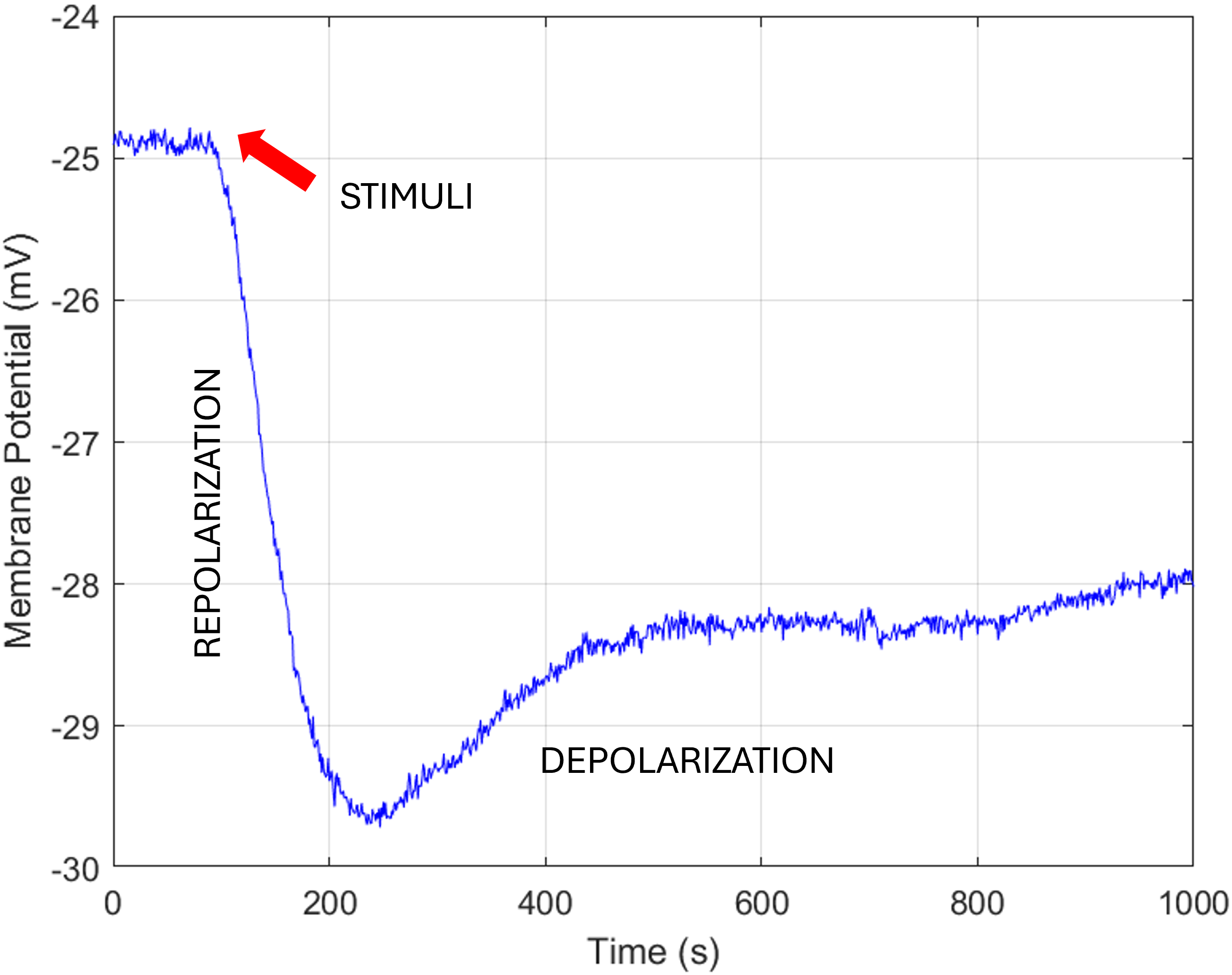}
    \caption{Dark-induced AP within a NETO response, recorded from tobacco at artificial light
    offset in the Agrowbox dataset.}
    \label{fig:NETO_ER_dark}
\end{figure}

Sudden light transitions elicit threshold-dependent APs with coupled light-dark responses~\cite{yao23, trebacz14, koselski08}, as shown in Figs.~\ref{fig:NETO_ER_light}
and~\ref{fig:NETO_ER_dark}, while gradual illumination generates APs characterized by prolonged $\mathrm{Ca}^{2+}$-mediated activity without dark-induced counterparts~\cite{tran19}, as in Fig.~\ref{fig:POCE_ER_light}. These findings, reported across species and experimental protocols~\cite{elzenga95, plieth98}, suggest that light onset velocity influences AP waveform morphology and functional coupling to dark transitions. The present work tests this hypothesis
under repeated photoperiodic conditions, defining NETO and POCE as distinct AP classes distinguished by their triggering dynamics, waveform characteristics, and dark-response coupling.

\subsection{Molecular Communication}
\label{sec:MolecularCommunication}
MC in plants spans from intracellular to plant-plant scales, coupling fast
electrical/$\mathrm{Ca}^{2+}$/ROS events to slower chemical transport that encodes, conveys, and decodes information for development, acclimation, and defense. In the encoder-channel-decoder view, local stimuli modulate biosynthesis and release of information molecules (e.g., phytohormones, peptides, RNAs) that traverse channels with dispersion, delay,
and reactive losses, and are decoded by receptor networks that map temporal-concentration patterns into gene regulatory programs~\cite{magarini20}. Plant MC channels are
well-approximated by advection-diffusion-reaction formalisms supporting communication-theoretic constructs such as capacity bounds, mutual information, and modulation/detection
design~\cite{kuran21}.

From a modulation/detection perspective, plant MC supports concentration-, type-, timing-, and
spatial-domain encodings. In vascular channels, concentration or type modulation (e.g., ABA
versus JA) is decoded by adaptive-threshold detectors accounting for inter-symbol interference (ISI)~\cite{kuran21, unluturk17}. Engineering proposals to harness genetically engineered bacteria as in situ sensors inside plant tissues offer a pragmatic route to augment native MC and create hybrid bio-cyber links for crop monitoring~\cite{magarini20}. 
This MC-centric abstraction provides a unifying language connecting molecular biology, plant physiology, and communication theory, guiding sensor placement, feature extraction, and decoder design for resilient plant-centered sensing systems~\cite{kuran21, magarini20}. The present work focuses on the encoding stage of this communication chain, characterizing how light stimuli are transduced into AP waveforms.

\subsection{AP Generation via HH-based Model}
\label{sec:HH}

This subsection introduces the classical HH model and its adaptation to plant electrophysiology,
providing the theoretical basis for the model developed in Section~\ref{sec:system_model}.

The HH model describes how APs are generated and propagated in animal excitable cells through
voltage-gated ion channels~\cite{abbott90}. The membrane potential $V_{\mathrm{m}}$ is
determined by the balance of ionic currents through channels for $\mathrm{Na}^+$, $\mathrm{K}^+$,
and leakage $\mathrm{L}$:
\begin{equation}
C_{\mathrm{m}} \frac{dV_{\mathrm{m}}}{dt} = I_{\mathrm{ext}} - \left( I_{\mathrm{Na}} +
I_{\mathrm{K}} + I_{\mathrm{L}} \right),
\label{eq:HH_membrane}
\end{equation}
where $C_{\mathrm{m}}$ is the membrane capacitance and each ionic current follows Ohm's law:
\begin{equation}
I_{i} = g_{i}(V_{\mathrm{m}} - E_{i}), \quad i \in \{\mathrm{Na}, \mathrm{K}, \mathrm{L}\},
\label{eq:ionic_current}
\end{equation}
where $g_i$ is the instantaneous conductance and $E_i$ the reversal potential. The conductances
are expressed as $g_{\mathrm{Na}} = \bar{g}_{\mathrm{Na}}\, m^3 h$ and
$g_{\mathrm{K}} = \bar{g}_{\mathrm{K}}\, n^4$, where $n$ governs $\mathrm{K}^+$ channel
activation, $m$ governs $\mathrm{Na}^+$ channel activation, and $h$ governs $\mathrm{Na}^+$
channel inactivation. Each gating variable obeys first-order kinetics~\cite{abbott90}:
\begin{equation}
\frac{dX}{dt} = \alpha_{X}(V_{\mathrm{m}})(1 - X) - \beta_{X}(V_{\mathrm{m}})\, X, \quad
X \in \{m, h, n\},
\label{eq:gating_kinetics}
\end{equation}
where $\alpha_X$ and $\beta_X$ are voltage-dependent forward and backward rate functions,
respectively. The exponents reflect cooperative gating: $\mathrm{K}^+$ channels require four
independent $n$-gates ($n^4$), while $\mathrm{Na}^+$ channels require three $m$-gates and one
$h$-gate ($m^3 h$)~\cite{abbott90}.

Despite its origins in animal neuroscience, the HH formalism has demonstrated remarkable applicability to plant APs~\cite{sukhov09, awan19}, with recent adaptations accommodating the significantly longer durations of plant APs compared to neuronal counterparts~\cite{awan19,
awan21}. The biophysical model by Yao et al.~\cite{yao23} for blue light-stimulated \textit{Arabidopsis thaliana} mesophyll cells, describing $\mathrm{Ca}^{2+}$, $\mathrm{K}^+$, $\mathrm{Cl}^-$, and $\mathrm{H}^+$ ionic currents through voltage-dependent gating kinetics, constitutes the starting point of this work. Rather than retaining voltage-dependent rate
functions, constant rate parameters $\alpha$ and $\beta$ are adopted here to reduce parameter dimensionality and improve identifiability when fitting across multiple recording days, while retaining sufficient flexibility to reproduce key waveform features.

Let the gating kinetics with constant rates be:
\begin{equation}
\frac{dX}{dt} = \alpha_X (1 - X) - \beta_X X, \quad X \in \{n, m, h\}.
\label{eq:gating_plant}
\end{equation}
In this formulation for plant cells, $m$ governs $\mathrm{Ca}^{2+}$ channel activation, $h$
governs inactivation for $\mathrm{Cl}^{-}$ channels and activation for $\mathrm{H}^{+}$
channels, and $n$ governs $\mathrm{K}^{+}$ channel activation. The ionic currents are:
\begin{equation}
I_{\mathrm{K}} = \bar{g}_{\mathrm{K}}\, n^4 (V_{\mathrm{m}} - E_{\mathrm{K}}),
\end{equation}
\begin{equation}
I_{\mathrm{Ca}} = \bar{g}_{\mathrm{Ca}}\, m^3 (V_{\mathrm{m}} - E_{\mathrm{Ca}}),
\end{equation}
\begin{equation}
I_{\mathrm{Cl}} = \bar{g}_{\mathrm{Cl}}\, m^2 h\, (V_{\mathrm{m}} - E_{\mathrm{Cl}}),
\end{equation}
\begin{equation}
I_{\mathrm{H}} = \bar{g}_{\mathrm{H}}\, h\, (V_{\mathrm{m}} - E_{\mathrm{H}}),
\end{equation}
\begin{equation}
I_{\mathrm{L}} = \bar{g}_{\mathrm{L}}\, (V_{\mathrm{m}} - E_{\mathrm{L}}),
\end{equation}
and the membrane equation becomes:
\begin{equation}
C_{\mathrm{m}} \frac{dV_{\mathrm{m}}}{dt} = I(t) - I_{\mathrm{K}} - I_{\mathrm{Ca}} -
I_{\mathrm{Cl}} - I_{\mathrm{H}} - I_{\mathrm{L}},
\label{eq:v_m_final}
\end{equation}
where $\bar{g}_{i}$ denotes the maximum conductance of the corresponding ion channel. The
exponents reflect cooperative gating assumptions from classical HH theory and its extensions to
plant electrophysiology~\cite{sukhov09, yao23}. All parameter values are selected from the
literature, as summarized in Table~\ref{tab:parameters}.

\begin{figure*}[!t]
    \centering
    \includegraphics[width=1\linewidth]{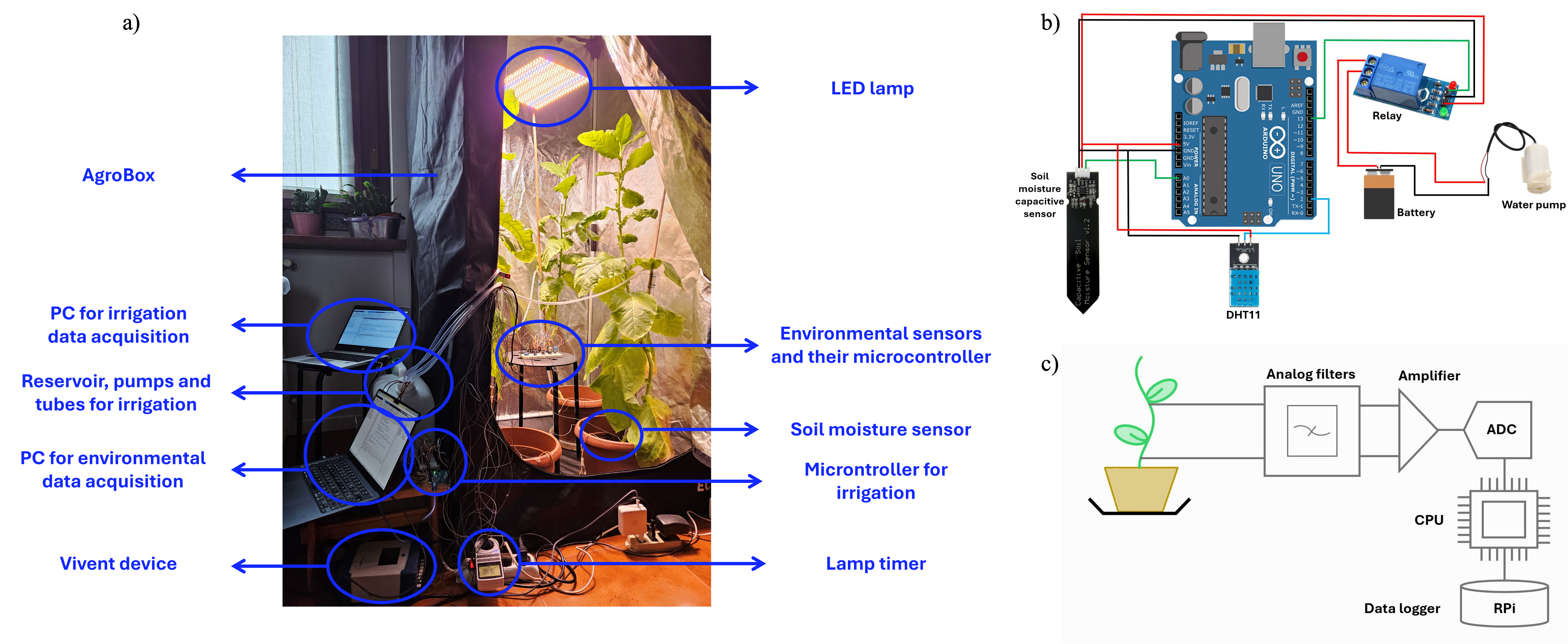}
    \begin{subfigure}[t]{0.32\linewidth}
        \centering
        \caption{}
        \label{SETUP}
    \end{subfigure}\hfill
    \begin{subfigure}[t]{0.32\linewidth}
        \centering
        \caption{}
        \label{ARDUINO}
    \end{subfigure}\hfill
    \begin{subfigure}[t]{0.32\linewidth}
        \centering
        \caption{}
        \label{ACQ}
    \end{subfigure}
    \caption{Agrowbox panel-specific descriptions. (a) Agrowbox composition. (b) Arduino-based control system: sensors and actuators. (c) Acquisition system.}
    \label{fig:setup_main}
\end{figure*}

\begin{table}[!t]
\renewcommand{\arraystretch}{1.4}
\caption{PARAMETERS FOR PLANT CELL $V_{\mathrm{m}}$}
\label{tab:parameters}
\centering
\begin{tabular}{@{}lcc@{}}
\hline
\textbf{Parameter} & \textbf{Value} & \textbf{Description} \\
\hline
$C_{\mathrm{m}}$ & $1 \, \mu\text{F/cm}^2$ & Membrane capacitance \\
$\bar{g}_{\mathrm{K}}$ & $44 \, \text{mS/cm}^2$~\cite{plieth98} & Max $\mathrm{K}^+$ conductance \\
$\bar{g}_{\mathrm{Ca}}$ & $183 \, \text{mS/cm}^2$~\cite{deangeli09} & Max $\mathrm{Ca}^{2+}$ channel conductance \\
$\bar{g}_{\mathrm{Cl}}$ & $31 \, \text{mS/cm}^2$~\cite{lew91} & Max $\mathrm{Cl}^-$ channel conductance \\
$\bar{g}_{\mathrm{H}}$ & $17 \, \text{mS/cm}^2$~\cite{yao23} & Max $\mathrm{H}^+$ channel conductance \\
$\bar{g}_{\mathrm{L}}$ & $21 \, \text{mS/cm}^2$~\cite{yao23} & Max leak channel conductance \\
$E_{\mathrm{K}}$ & $-65 \, \text{mV}$~\cite{plieth98} & $\mathrm{K}^+$ reversal potential \\
$E_{\mathrm{Ca}}$ & $43 \, \text{mV}$~\cite{deangeli09} & $\mathrm{Ca}^{2+}$ reversal potential \\
$E_{\mathrm{Cl}}$ & $-25 \, \text{mV}$~\cite{lew91} & $\mathrm{Cl}^-$ reversal potential \\
$E_{\mathrm{H}}$ & $-57.4 \, \text{mV}$~\cite{yao23} & $\mathrm{H}^+$ reversal potential \\
$E_{\mathrm{L}}$ & $9 \, \text{mV}$~\cite{yao23} & Leak reversal potential \\
\hline
\end{tabular}
\vspace{0.6mm}
\end{table}

\section{System Model}
\label{sec:system_model}

To characterize plant electrophysiological responses under controlled photoperiod conditions,
a custom growth chamber was designed to ensure repeatable light transitions and stable environmental parameters throughout the recording period.

\subsection{Agrowbox Design}
\label{sec:agrowbox}
The growbox (herein termed "AgroBox") in Fig.~\ref{SETUP} measured $1.2\,\text{m}$$\,\times \,$$1.2\,\text{m} $$\,\times\, $$2\,\text{m}$ and housed four tobacco specimens. It featured a $700\,$W LED lamp for plant growth, connected to a timer to maintain a consistent 12-hour day-night cycle, switching on at 7:00AM and off at 7:00PM. The lamp was positioned $40$-$50\,$cm above the plants for optimal light exposure.
An Arduino UNO microcontroller monitored environmental conditions. The DHT11 sensor continuously recorded ambient temperature and humidity, while an automated irrigation system regulated water delivery based on real-time soil moisture readings from capacitive sensors placed in each pot (see Fig.~\ref{ARDUINO}). These sensors communicated with the Arduino, activating $5\,$V submersible pumps via a 4-channel relay when moisture dropped below a predefined threshold. The pumps drew water from a $20\,$L reservoir through plastic tubing. Each activation lasted six seconds, ensuring adequate watering. With dedicated sensors and pumps for each plant, irrigation was individually regulated. The optimal moisture threshold was set at $75\%$, determined through calibration, with $0\%$ indicating dry soil and $100\%$ indicating saturation~\cite{bekkari24}.

\subsection{Biological Aspect}
\label{sec:bio}
Tobacco was selected due to its established role in plant biology research and well-documented electrophysiological responses~\cite{van93}. Its robust morphology, including large leaves and sturdy stems, facilitates bioelectrical measurements~\cite{bekkari24}. Finally, its extensive genetic and physiological characterization, combined with economic significance, establishes tobacco as an optimal model organism to investigate plant electrophysiology and its underlying molecular mechanisms.

\subsection{Electrophysiological Data Acquisition}
\label{sec:acquisition}
Electrophysiological measurements from tobacco plants were acquired using the Vivent biosignal amplifier (see Fig.~\ref{ACQ}). For each plant, a dual-electrode configuration was adopted, with both electrodes inserted into the stem to capture extracellular potential differences. The electrodes were connected to a DC-coupled amplifier, and the amplified signals were digitized via an analog-to-digital converter operating at a sampling frequency of $f_s $$\,= \,$$1\,$Hz. The processed digital signals were logged and stored on a Raspberry Pi (RPi), which served as the data logger.
\begin{figure*}[!t]
    \centering
    \includegraphics[width=\linewidth]{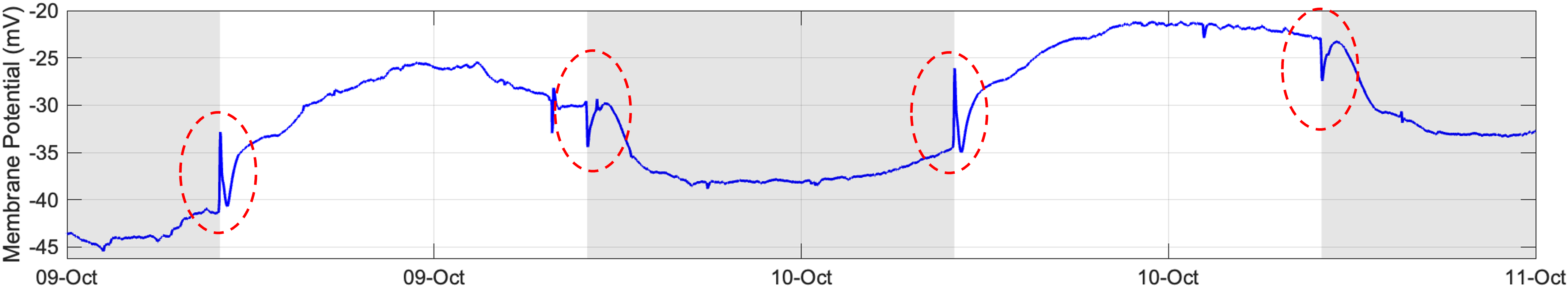}
    \caption{NETO responses recorded from tobacco over the recording period. Grey shaded regions indicate dark periods (light-off), while unshaded regions correspond to light-on conditions. Red dashed circles highlight the light-induced and dark-induced APs.}
    \label{fig:NETO_3days}
\end{figure*}

\begin{figure*}[!t]
    \centering
    \includegraphics[width=\linewidth]{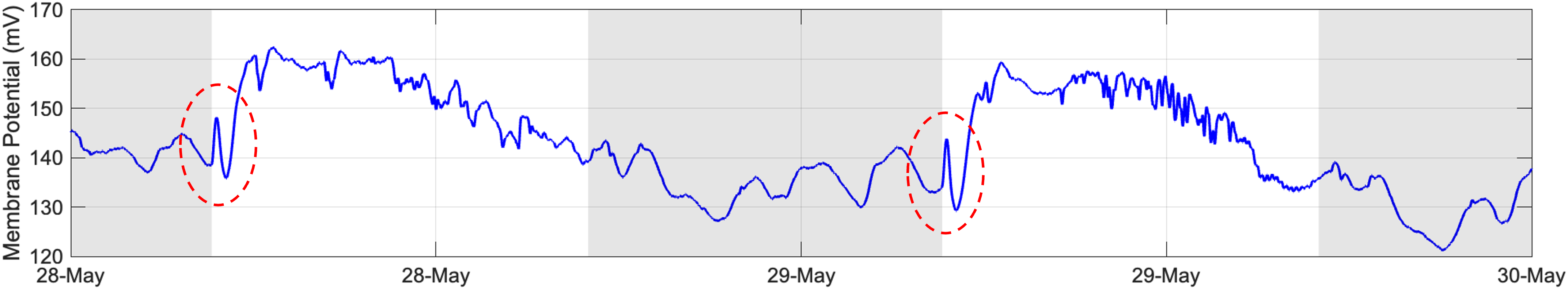}
    \caption{POCE responses recorded from tomato, obtained from the external Vivent SA dataset~\cite{vivent}. Grey shaded regions indicate dark periods (light-off), while unshaded regions correspond to light-on conditions.}
    \label{fig:POCE_3days}
\end{figure*}

\section{Data Analysis}
\label{sec:data_analysis}

Recordings from both datasets span three consecutive days to assess reproducibility. For the acquired dataset, each day of data is divided into a 45-minute morning session beginning at 6:55AM to capture light-induced APs, and a 15-minute evening session from 6:55PM to record dark-induced APs (Fig.~\ref{fig:NETO_3days}). To enhance signal clarity and mitigate noise, a low-pass filter at $30\,$Hz and notch filters at $50\,$Hz and $100\,$Hz were applied. The recorded signals were then resampled to a frequency of $f_s$$\,= \,$$1\,$Hz.
For each session, two features are extracted: peak amplitude, defined as the maximum voltage deviation from the resting potential during depolarization, and total duration, measured as the time interval from the onset of depolarization to the return to baseline following repolarization~\cite{garcia21, mudrilov21}. Together, these features capture the basic dynamics of depolarization and repolarization phases. Reproducibility is evaluated by computing the mean and standard deviation for each feature across the recording days.

\begin{figure*}[t]
    \centering
    \begin{minipage}{0.32\textwidth}
        \centering
        \includegraphics[width=\linewidth]{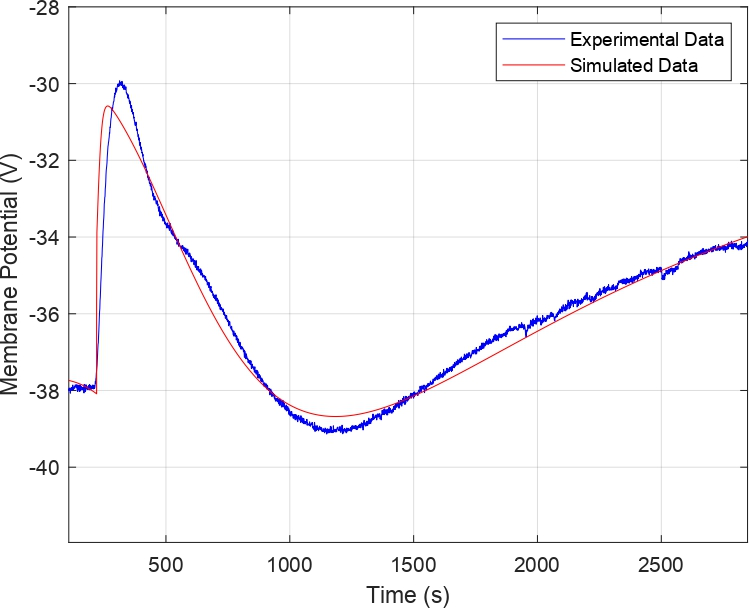}
    \end{minipage}\hfill
    \begin{minipage}{0.32\textwidth}
        \centering
        \includegraphics[width=\linewidth]{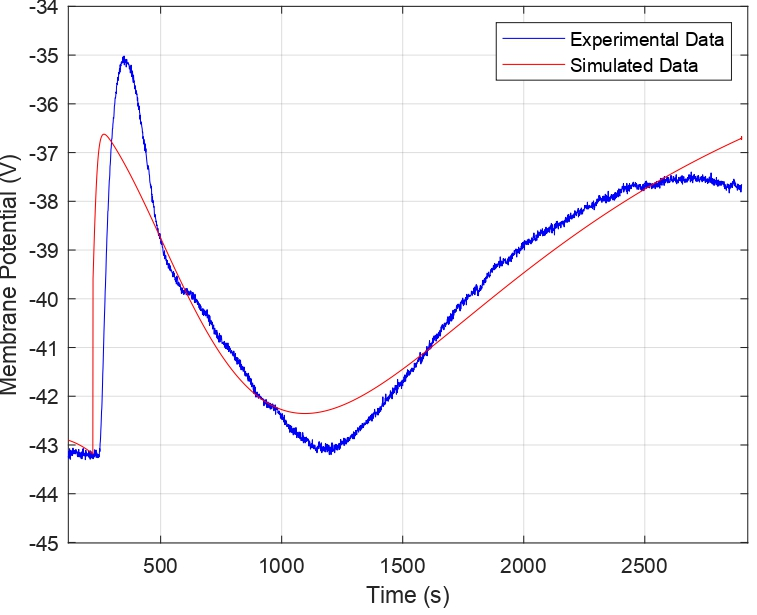}
    \end{minipage}\hfill
    \begin{minipage}{0.32\textwidth}
        \centering
        \includegraphics[width=\linewidth]{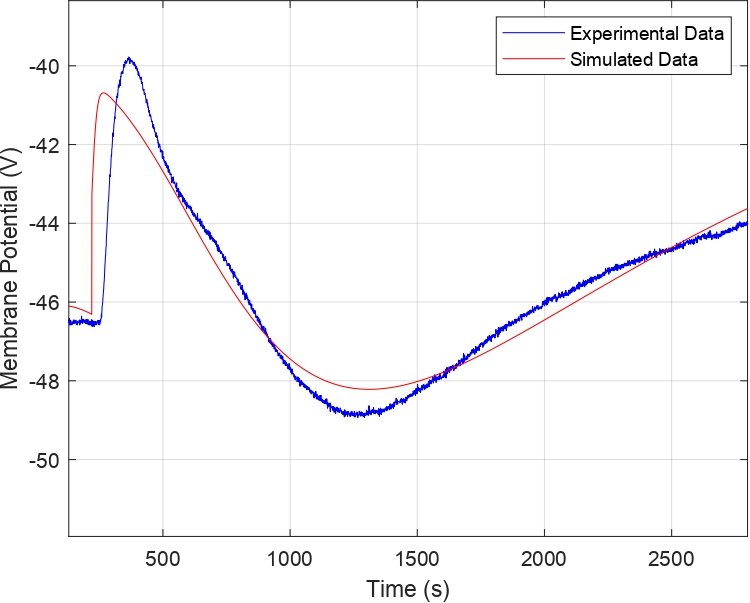}
    \end{minipage}
    \caption{Light-Induced NETO APs: comparison between measured data and model-obtained data.}
    \label{fig:APAM}
\end{figure*}
\begin{figure*}[t]
    \centering
    \begin{minipage}{0.32\textwidth}
        \centering
        \includegraphics[width=\linewidth]{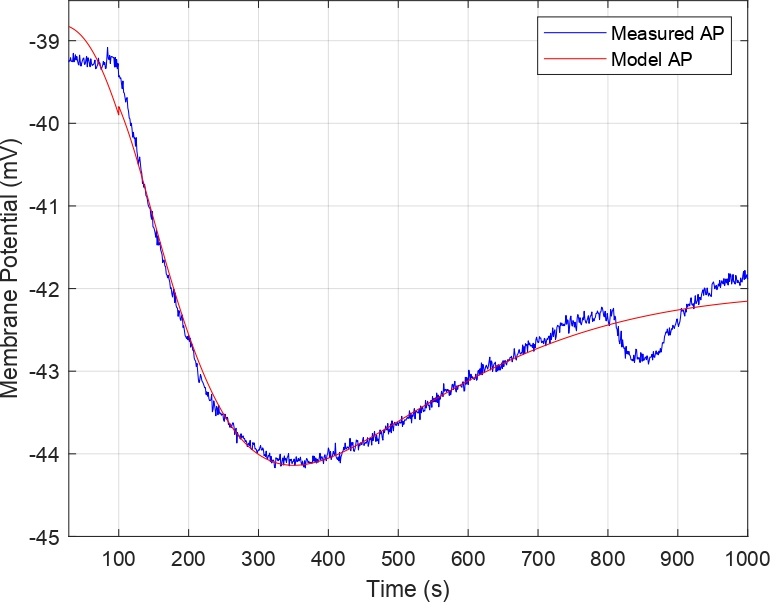}
    \end{minipage}\hfill
    \begin{minipage}{0.32\textwidth}
        \centering
        \includegraphics[width=\linewidth]{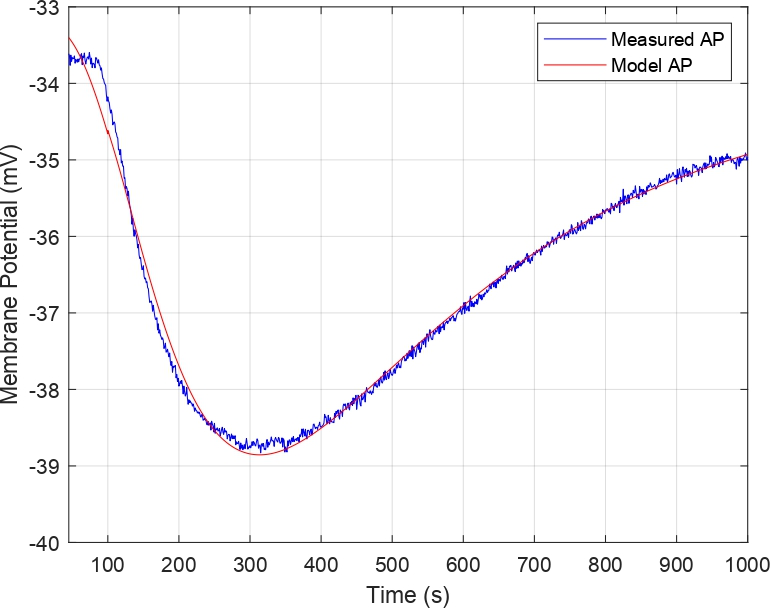}
    \end{minipage}\hfill
    \begin{minipage}{0.32\textwidth}
        \centering
        \includegraphics[width=\linewidth]{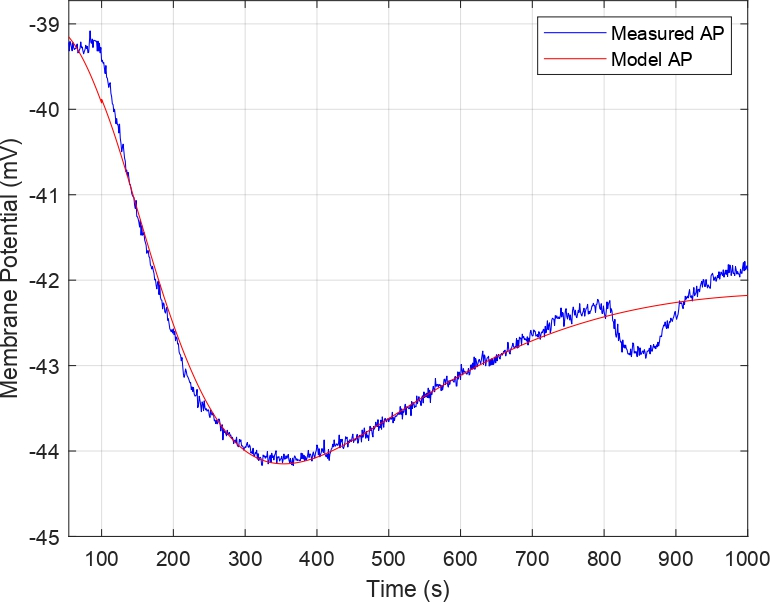}
    \end{minipage}
    \caption{Dark-Induced NETO APs: comparison between measured data and model-obtained data.}
    \label{fig:APAN}
\end{figure*}
\begin{figure*}[t]
    \centering
    \begin{minipage}{0.32\textwidth}
        \centering
        \includegraphics[width=\linewidth]{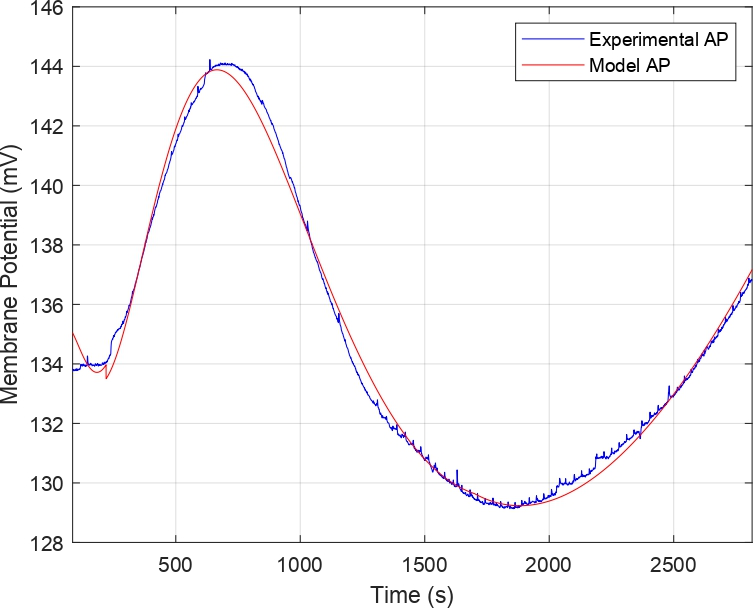}
    \end{minipage}\hfill
    \begin{minipage}{0.32\textwidth}
        \centering
        \includegraphics[width=\linewidth]{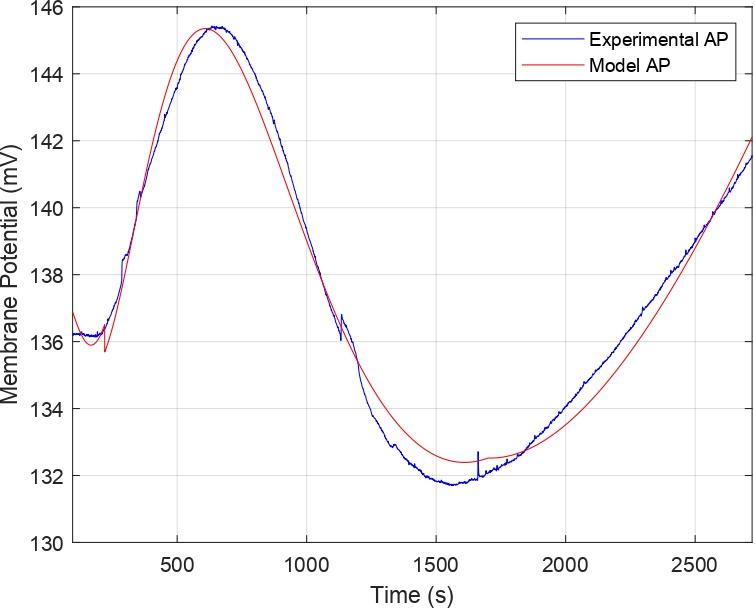}
    \end{minipage}\hfill
    \begin{minipage}{0.32\textwidth}
        \centering
        \includegraphics[width=\linewidth]{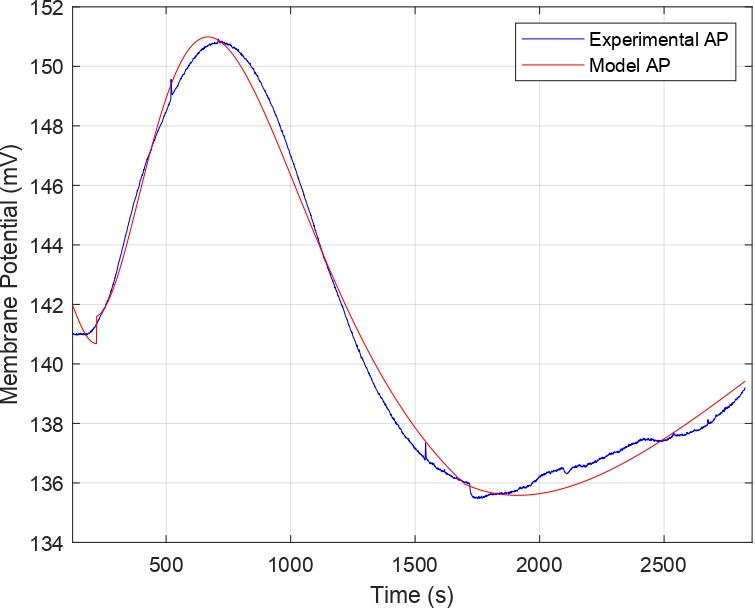}
    \end{minipage}
    \caption{Light-Induced POCE APs: comparison between measured data and model-obtained data.}
    \label{fig:APBM}
\end{figure*}

\subsection{External Dataset}
The external dataset was provided by Vivent SA and involves tomato plants grown in a greenhouse under natural light supplemented with reinforced lighting as needed~\cite{vivent}. For each plant, a dual-electrode configuration was adopted, with both electrodes inserted into the stem.
Signals were sampled at $1\,$Hz and stored locally. Three consecutive days of recordings were selected to match the analysis period of the acquired dataset (Fig.~\ref{fig:POCE_3days}).
Further experimental details regarding the Vivent dataset, including specific growth conditions and measurement protocols, are proprietary and not publicly disclosed by the data provider~\cite{vivent}. Its inclusion serves as a first validation of the proposed model across different species and illumination conditions, with POCE responses being exclusive to natural photoperiods not reproducible in the Agrowbox.

\section{Results}
\label{sec:results}

In the following, we present the electrophysiological recordings obtained under artificial and natural illumination, and evaluate the ability of the HH-based model to reproduce the observed AP waveforms under both NETO and POCE conditions.

\subsection{Experimental Observations}
\label{sec:obs}
Electrophysiological recordings revealed distinct plant responses to photo-stimuli. As shown in Fig.~\ref{fig:NETO_3days}, tobacco under controlled artificial illumination exhibited a consistent AP-coupled pattern, with light- and dark-induced APs synchronized at 7:00~AM and 7:00~PM within the 24-hour cycle. This temporal consistency supports the hypothesis that abrupt light transitions elicit coupled light-dark AP responses, here termed NETO~\cite{yao23, trebacz14, koselski08}.
Fig.~\ref{fig:POCE_3days} presents the POCE phenomenon observed in tomatoes. These recordings show light-induced APs occurring exclusively during sunrise. This asymmetric pattern is consistent with literature evidence indicating that gradual illumination changes generate prolonged responses without dark-induced counterparts~\cite{tran19, koselski08}.
The comparative assessment reveals differences between NETO and POCE.  
POCE responses exhibit larger peak amplitudes ($10.73 \pm 2.21$~mV) compared to NETO ($7.85 \pm 1.36$~mV), and longer durations ($41.33 \pm 4.92$~min vs. $26.25 \pm 1.50$~min). NETO dark-induced APs display smaller amplitudes and shorter durations than light-induced APs.

\subsection{Model Fitting}
 \label{sec:fit}

The HH-based model defined by Eqs.~\ref{eq:gating_plant}-\ref{eq:v_m_final} was fitted to experimental AP waveforms. The light stimulus enters the model through the external current term $I(t)$ in Eq.~\ref{eq:v_m_final}, represented as a step function that activates at stimulus onset with amplitude $I_0$ and time constant $\tau_I$. For NETO, $\tau_I$ was set to a small value, while for POCE, $\tau_I$ was larger to reflect the gradual sunrise.
Conductances and reversal potentials were fixed as defined in Table~\ref{tab:parameters}. The fitted parameters were the gating rate constants ($\alpha_m$, $\beta_m$, $\alpha_h$, $\beta_h$, $\alpha_n$, $\beta_n$) and the stimulus. Parameter estimation was performed using constrained nonlinear optimization minimizing the sum of squared errors between model output and experimental data. The differential equations were integrated using forward Euler with a time step of 1ms, and model output was interpolated to the 1~Hz experimental sampling rate for error computation.

\subsection{Model Performace}
\label{sec:perf}

Figs.~\ref{fig:APAM} and \ref{fig:APAN} compare measured and model-generated NETO APs for light and dark transitions, respectively. The model captures key features including resting potential, depolarization rate, peak amplitude, and duration. For light-induced NETO APs (Fig.~\ref{fig:APAM}), the model reproduces the sharp depolarization and peak, though it predicts slightly accelerated repolarization in some cases. Dark-induced NETO APs (Fig.~\ref{fig:APAN}) show strong agreement across depolarization, peak, repolarization, and hyperpolarization phases.

Fig.~\ref{fig:APBM} presents light-induced POCE APs. The external dataset provided by Vivent SA ~\cite{vivent}. APs corresponding to sunrise were extracted for fitting. The model captures the slower rise time and extended duration characteristic of POCE. The constant-rate gating kinetics reproduce depolarization and repolarization without voltage-dependent rate functions, consistent with the slower stimulus dynamics under natural light conditions.

\section{Discussion}
\label{sec:discussion}

The results are interpreted in light of the stimulus velocity hypothesis, examining how illumination dynamics rather than species-specific factors drive the observed differences in AP morphology between NETO and POCE conditions.
The light-induced APs recorded under artificial and natural illumination exhibited different characteristics despite both representing responses to increasing light intensity. NETO light-induced APs displayed sharper depolarization, smaller amplitude, and shorter duration compared to POCE responses, suggesting that the temporal profile of the light stimulus shapes the resulting waveform morphology~\cite{koselski08, yao23, tran19}. Abrupt light onset may activate ion channels rapidly and synchronously, producing a brief coordinated response, whereas gradual illumination could recruit channels progressively, sustaining $\mathrm{Ca}^{2+}$ influx and prolonging depolarization~\cite{plieth98, tran19}. The presence of dark-induced APs only in the NETO condition further indicates that rapid transitions establish membrane states conducive to subsequent dark-triggered responses, a coupling absent when light onset occurs slowly~\cite{koselski08}. The constant-rate gating simplification proved sufficient to capture AP waveforms across both conditions. Voltage-dependent formulations~\cite{yao23} would provide mechanistic insight into channel kinetics but require higher temporal resolution.
The reproducibility observed across recording period indicates potential for using light-induced APs as non-invasive indicators of environmental conditions or plant status. From an MC perspective, NETO and POCE represent distinct encoding schemes where waveform characteristics convey information about stimulus dynamics~\cite{magarini20}.

\section{Conclusion}
\label{sec:conclusion}

This study characterized light-induced APs under artificial (NETO) and natural (POCE) illumination conditions and demonstrated that an HH-based model with constant-rate gating parameters can reproduce the observed waveform diversity. The systematic differences in depolarization rate, amplitude, and duration between NETO and POCE responses suggest that stimulus velocity influences AP morphology, with abrupt transitions producing coupled light-dark responses absent under gradual illumination.
The reproducibility observed across recording days supports the potential use of light-induced APs as non-invasive indicators of plant status. 
Future work will employ controlled variation of light onset rates within a single species to directly test the stimulus velocity hypothesis, followed by extension of the modeling framework to encompass the complete MC chain including signal propagation and physiological decoding.

\section*{Acknowledgments}
We would like to thank Nigel Wallbridge
(Vivent SA, CH) for giving us access to their data.

\bibliographystyle{IEEEtran} 
\bibliography{references.bib} 

\end{document}